# Adaptive conditional latent diffusion maps beam loss to 2D phase space projections

Alexander Scheinker[1,*] and Alan Williams[1]

[1]Applied Electrodynamics Group, Los Alamos National Laboratory, Los Alamos, New Mexico 87545, USA.
*ascheink@lanl.gov

Beam loss (BLM) and beam current monitors (BCM) are ubiquitous at particle accelerator around the world. These simple devices provide non-invasive high level beam measurements, but give no insight into the detailed 6D $(x,y,z,p_x,p_y,p_z)$ beam phase space distributions or dynamics. We show that generative conditional latent diffusion models can learn intricate patterns to map waveforms of tens of BLMs or BCMs along an accelerator to detailed 2D projections of a charged particle beam's 6D phase space density. This transformational method can be used at any particle accelerator to transform simple non-invasive devices into detailed beam phase space diagnostics. We demonstrate this concept via multi-particle simulations of the high intensity beam in the kilometer-long LANSCE linear proton accelerator.

## I. INTRODUCTION

Control of the 6D $(x,y,z,p_x,p_y,p_z)$ phase space distribution of beams is important for all particle accelerators over a wide span of applications including beam- based imaging for material science, accelerator-based light sources, and plasma wakefield-based accelerators.

At all large high intensity beam particle accelerators, such as the Los Alamos Neutron Science Center (LAN-SCE) [1] or the Spallation Neutron Source (SNS) [2], the beam's phase distribution must be controlled for proper acceleration and to prevent beam loss by matching to the accelerator's magnetic focusing lattice. At particle accelerator-based free electron laser (FEL) light sources, such as the Linac Coherent Light Source [3], the European X-ray FEL [4], and the Swiss FEL [5] the beam's phase space defines the properties of the generated light and must be adjusted between different experiments. For plasma wakefield acceleration (PWA), such as the Advanced Proton Driven Plasma Wakefield Acceleration Experiment (AWAKE) at a CERN [6], or the Facility for Advanced Accelerator Experimental Tests (FACET and FACET-II) at SLAC [7, 8] the details of the beam's 6D phase space density influence the PWA process. For laser-driven PWA [9], controlling the quality of the created beam's 6D phase space is challenging. Non- destructive views of a beam's 6D phase space for PWA has the potential to help in the understanding and optimization of complex PWA processes.

When detailed beam phase space measurements are available advanced control theory-based adaptive machine learning (ML) techniques have been developed to quickly automatically shape the beam's phase space [10]. For non-invasive beam measurements, beam loss (BLM) and beam current monitors (BCM) are ubiquitous, pro- viding a measure of beam quality, but give no insight into a beam's detailed 6D phase space density. They simply measure 1D signals proportional to the total number of particles at a given accelerator location [11].

Recently, the world's first 6D beam measurement was conducted at the SNS Beam Test Facility, but this re- quired 32 hours (now reduced to 18 hours) and 5675740 measurements [12]. ML-based methods have the potential to significantly speed up such measurements for real-time control. There are many efforts underway to develop ML-based tools as virtual beam diagnostics including adaptive latent space tuning of generative autoencoders for tracking time-varying beams [13, 14], combining generative ML models with physics models [15–17], latent evolution models [18], and normalizing flows [19].

In the ML/AI community, diffusion-based models are the state-of-the-art for generative accurate representations of high dimensional complex data [20–25]. Recently, the first application of diffusion was demonstrated as a virtual diagnostic to generate megapixel resolution virtual views of the (z,E) longitudinal phase space of the electron beam in the European X-ray FEL [26]. That method was then generalized to an adaptively tuned conditional diffusion approach for generating high resolution representations of all of the unique 2D projections of the HiRES UED beam's 6D phase space distribution [27]. A physics-constrained super-resolution diffusion- based method has been developed in which a beam's en- tire 6D phase space density is generated as a 6D tensor from which 2D views are projected [28].

In this paper, we demonstrate that conditional generative latent diffusion can find subtle patterns in waveforms of tens of BLM or BCM measurements along the length of an accelerator to map these vectors to highly detailed representations of the various 2D projections of a beam's 6D phase space distribution, as shown in Figure 2. This transformational result provides a method for transforming simple non-invasive BLM and BCM devices into detailed beam phase space diagnostics. This same approach can be applied at any particle, completely transforming the amount of non-invasive beam diagnostics information that is available for tuning and setup. As a proof of principal, we demonstrate this generative AI concept via multi-particle simulations of high intensity beams in the kilometer-long LANSCE linear proton accelerator as shown in Figure 1.

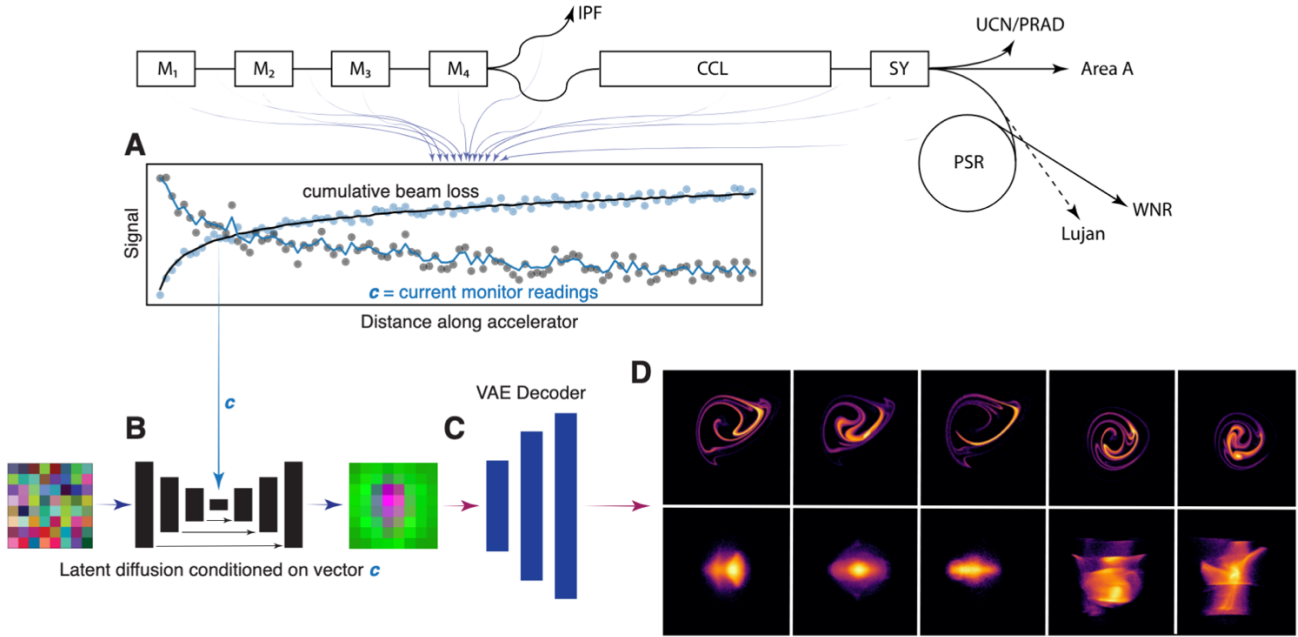

FIG. 1. Overview of the latent diffusion approach for mapping beam current or loss monitor readings to 2D phase space projections. **A**: Non-invasive beam current or loss monitor readings are recorded along the length of the accelerator. **B**: The vector $\mathbf{c} = (\mathbf{c_1}, \ldots, \mathbf{c_n})$ of beam current at various accelerator locations together with accelerator parameters is used to condition the diffusion process. The conditional diffusion process generates a latent embedding representation of the beam's 2D phase space projection. **C**: The decoder of a variational autoencoder generates the desired 2D phase space projection from its latent representation. **D**: Various examples of 2D beam projections are shown. In this setup we can generate any one of the 15 unique 2D projections of the beam's phase space and here various examples of the $(z, E)$ and $(x, y)$ projections are shown.

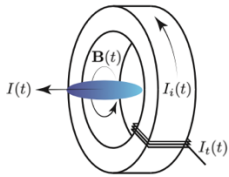

FIG. 2. In an inductive beam current monitor, the time-varying flux of a passing beam's magnetic field $\mathbf{B}(t)$ induces a current $I_i(t)$ in a conductor, which is measured as a current $I_t(t)$ proportional to the beam's current $I(t)$.

Figure 2 shows an overview of an inductive BCM whose operation is described by the Maxwell–Faraday equation

$$\oint_{\partial \Sigma} \mathbf{E} \cdot d\mathbf{l} = -\int_{\Sigma} \frac{\partial \mathbf{B}}{\partial t} \cdot d\mathbf{A}, \quad \mathbf{B}(r,t) = \frac{\mu_0 I(t)}{2\pi r} \hat{\theta}, \quad (1)$$

where I(t) is beam current, r is transverse distance from the beam center, $\mathbf{B}(r,t)$ is the beam's magnetic field, and $\mathbf{E}$ is the electric field induced by the beam passing through a conducting loop. The induced electric field creates a current in the loop which is then coupled through a simple AC transformer to a current measurement pickup, giving a signal proportional to beam current.

## II. DATA

For this work we utilize the particle tracking code HP-Sim which includes 3D space charge and is implemented in CUDA to run extremely fast on a GPU [29]. The code simulates the transport of a charged particle beam from the source to the end of the coupled cavity linac (CCL) section of the LANSCE accelerator through approximately 1 km as it traverses 48 resonant frequency (RF) accelerating cavities. The data is generated by setting an initial beam condition and then randomly perturbing the amplitude and phase of the first 4 RF cavities of LANSCE within uniform distributions over a range of ±3 degrees and ±3 % of the RF phase and amplitude setpoint, respectively, relative to a design tune of the RF system. Note that this is a significant change as operators typically adjust any one of these parameters at most ±1 degrees or ±1 % to maintain minimal beam loss while the system slowly varies with time due to factors such as temperature changes. For a fixed initial beam condition, 500 such simulations are run, each with $10^6$ macro-particles, and for each simulation 2D histograms are made for each of the unique 2D projections of the beam's 6D phase space at the exit of every single RF module over a 256 × 256 pixel grid. This process is repeated for 4 different initial conditions resulting in a data set of 2000 simulations where each simulation results in 15 images of 256×256 pixels each at 48 module locations.

The final dataset is 2000 × 48 × 15 = 1.44 million high resolution 256 × 256 pixel images. Of these simulations, from the 4 different runs having the same initial beam conditions, the training data consists of 400 simulations from each of the first 3 runs. The test set consists of 100 simulations from each of the first 3 runs and all 500 simulations of the 4th run. This split allows us to test the model's predictive ability relative to a fixed initial condition and varying RF parameters as well as a completely unseen initial condition together with varying RF parameters. The reasoning behind this dataset is that in practice the beam's initial conditions change on a slower time scale than the RF parameters. Every time that the beam source is replaced, the beam's initial conditions experience a step change and then while that source is run and the beam's initial conditions are changing very slowly, the entire RF system's characteristics continuously drift (on time scales of minutes-hours), requiring continuous manual re-tuning by the operators and beam physicists.

At each module m ∈ {1,2,...,48}, the 15 unique beam projections of the beam's 6D phase space density $\rho^m(x,y,z,p_x,p_y,p_z)$ are:

$$\rho_1^m(x,y), \ \rho_2^m(x,z), \ \rho_3^m(x,p_x), \ \rho_4^m(x,p_y), \ \rho_5^m(x,p_z)$$
$$\rho_6^m(y,z), \ \rho_7^m(y,p_x), \ \rho_8^m(y,p_y), \ \rho_9^m(y,p_z)$$
$$\rho_{10}^m(z,p_x), \ \rho_{11}^m(z,p_y), \ \rho_{12}^m(z,p_z)$$
$$\rho_{13}^m(p_x,p_y), \ \rho_{14}^m(p_x,p_z)$$
$$\rho_{15}^m(p_y,p_z)$$

We calculate a number proportional to the surviving beam current at each module by simply summing over the $\rho^m{}_{12}(z,p_z)$ projection

$$b_m = \int_z \int_{p_z} \rho^m(z, p_z) dp_z dz. \qquad (2)$$

Note that any projection can be used, they should all give identical answers. In practice this one was chosen be- cause the projection bounds were carefully selected along the modules to capture as much beam as possible for a wide range of RF settings while not making the bounds too wide so that details of the beam would be missed. Be- cause of this sometimes the edges of the beam are clipped by the size of the projection window, but this is rare and has negligible effects on the overall analysis.

In high energy particle accelerators the beam's longitudinal momentum typically dominates the other com- ponents with $p_z \gg p_x, p_y$, therefore instead of $p_x$ and $p_y$ we typically consider the quantities $x' = p_x/p_z$ and $y' = p_y/p_z$ which provide a measure of the beam's diver- gence from straight line motion. Furthermore, because of the resonant RF structures used to accelerate the beam, instead of considering the global z position of particles in the accelerator, it is also common to consider the phase φ of a particle in a bunch relative to a design particle, where φ is the phase shift of the RF accelerating field. In particular, for a given RF cavity amplitude set point $A_S$ and phase set point $\theta_S$, HPSim considers a design particle that arrives when the RF field's phase is at $\theta_S$, so that the electric field in the cavity is given by

$$E = A_s \sin(\omega t + \theta_s), \qquad (3)$$

then the "phase" of any particle in the bunch is defined by considering the distance Δz between the particle and the design particle, considering the particle velocity v at the entrance to the RF cavity, and calculating the amount of phase shift of the RF field between the two particles according to

$$\Delta\theta = \omega \Delta t = \omega \frac{v}{\Delta z}, \qquad (4)$$

and assigning a phase θ = $\theta_S$ +Δθ. For the 2D projection of the bunch at the RF cavity, the absolute phase $\theta_S$ is not important, and so we consider just the relative phase φ = Δθ to get the total phase spread of the bunch. One final change is that we typically consider the total kinetic energy of the beam, E, assuming that it is almost entirely due to the $p_z$ momentum component.

With these new phase space coordinates the 15 projections that we typically care about are the 15 unique beam projections of the beam's 6D phase space density $\rho^m(x, y, z, x', y', E)$ given by:

$$\rho_1^m(x,y) \ \rho_2^m(x,\varphi) \ \rho_3^m(x,x') \ \rho_4^m(x,y') \ \rho_5^m(x,E)$$
$$\rho_6^m(y,\varphi) \ \rho_7^m(y,x') \ \rho_8^m(y,y') \ \rho_9^m(y,E)$$
$$\rho_{10}^m(\varphi,x') \ \rho_{11}^m(\varphi,y') \ \rho_{12}^m(\varphi,E)$$
$$\rho_{13}^m(x',y') \ \rho_{14}^m(x',E)$$
$$\rho_{15}^m(y',E)$$

represented as a 256 × 256 pixel images. For example, near the front end, the Los Alamos Neutron Science Center (LANSCE) linear proton accelerator uses 100 MeV proton beams to create medical isotopes, for which the transverse beam size $\sigma_r$, average beam cur- rent, and beam energy are important as the beam is impacted onto a target for which the beam's kinetic energy controls the nuclear process of isotope creation and the beam must be rastered across the sample based on its size and current to prevent over-heating [30]. Near the end of LANSCE, an 800 MeV proton beam is impacted onto a tungsten target producing a high neutron flux for neutron diffraction for material science studies, for which only the average beam energy is important. At the end of LANSCE, proton radiography requires a precisely controlled longitudinal phase space (z, $p_z$) distribution of beam energy and bunch length $\sigma_z$ for beam- based dynamic imaging [31]. Throughout the LANSCE accelerator the beam's transverse (x,y) and ($p_x$, $p_y$) distributions must be controlled to keep the beam matched to the quadrupole magnet-based focusing lattice so that the beam's space charge does not rip it apart causing it to impact the accelerator's beam pipe.

### III. Variational Autoencoder

Our eventual goal is to create a map from RF settings and beam current measurements to the various 2D phase space projections described above via a conditional guided latent diffusion approach. The first step of latent diffusion is to compress a large 256x256 pixel image $\rho^m{}_j$ down to a much lower resolution latent representation $z^m{}_j$ to speed up the diffusion process that will then work directly in the latent space.

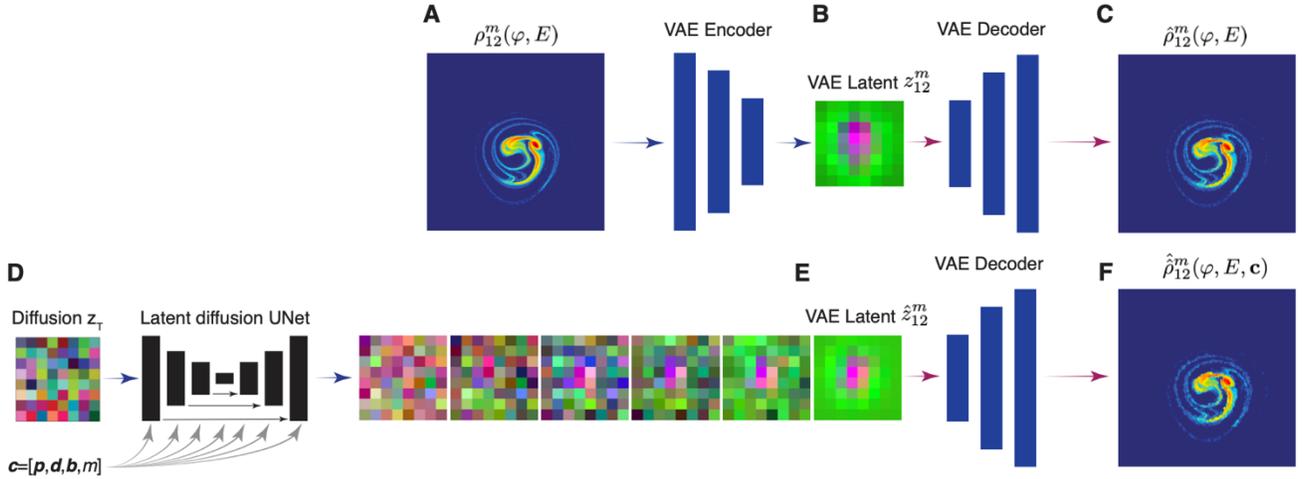

FIG. 3. Combining a variational autoencoder (VAE) with a diffusion model for latent diffusion. **A**: The VAE is used to compress a projection $\rho_j^m \in \mathbb{R}^{256\times 256}$ down to a much lower dimensional latent embedding $z_j^m \in \mathbb{R}^{8\times 8\times 3}$ (**B**) from which an estimate of the image is then reconstructed as $\hat{\rho}_j^m \in \mathbb{R}^{256\times 256}$ (**C**). **D**: A latent diffusion model, guided by a conditional vector **c** containing beam current measurements, is then trained to generate estimates of the latent embeddings $\hat{z}_j^m \in \mathbb{R}^{8\times 8\times 3}$ (**E**). **F**: The generated latent embeddings $\hat{z}_j^m \in \mathbb{R}^{8\times 8\times 3}$ are then passed back into the decoder of the VAE resulting in conditional generation of the high-resolution phase space projections $\hat{\hat{\rho}}_j^m \in \mathbb{R}^{256\times 256}$.

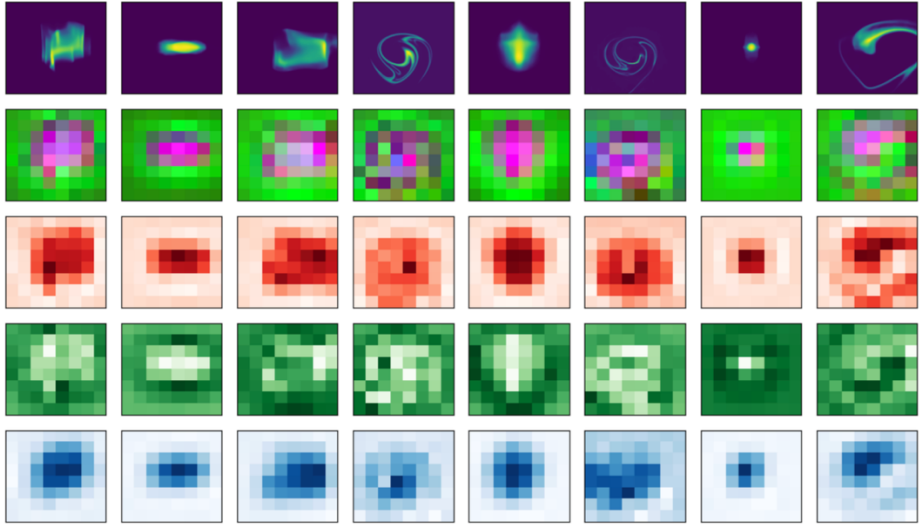

FIG. 4. The top row shows 8 random examples of 256 × 256 pixel 2D phase space projections of the beam. The second row shows the associated 8 × 8 pixel red, green, blue (RGB) 3-channel latent embedding images of the VAE. The third through fifth rows show the separate RGB channels of the latent embeddings. The red channel seems to be activated by the brightest regions of the original image and its interior, the green channel is highly active at the image boundaries, and the blue channel looks like a down sampled 8 × 8 pixel version of the original image, uniformly covering the support of the image.

For this dimensionality reduction task we first train a variational autoencoder (VAE) convolutional neural network which embeds the 2D phase space projections $\rho_j^{m_j} \in \mathbb{R}^{256\times 256}$ into latent representations $z_j^{m_j} \in \mathbb{R}^{8\times 8\times 3}$, from which estimates of the projections $\hat{\rho}_j^{m_j} \in \mathbb{R}^{256\times 256}$ are then reconstructed, as shown in Figure 3A. By embedding a 256×256 = 65536 pixel image into a 3-channel 8 × 8 × 3 = 192 pixel image, the data is compressed by a factor of 341. Using additional channels allows the VAE's latent representation to take into account various image properties such as edges, as shown in Figure 4. A high-level overview of the VAE is shown in Figure 5, demonstrating how much smaller the latent embedding is for fast latent diffusion. The resolution decreasing layers shown in Figure 5 each consist of a group of three 2D convolution layers with 3 × 3 filters and swish activation functions [32] followed by batch normalization [33]. The first group of 2D convolutions has 64 filters each. After each such set of three, the image resolution is reduced by a factor of 2 using a MaxPool 2D layer with a 2×2 pool size. This procedure is repeated until the images are reduced to 8 × 8 with the number of filters doubling at each stage. The network then splits into two branches, each of size 8 × 8 × 3, with one defining a mean image and the other defining a log variance. Together these define a 8 × 8 × 3 normal distribution from which samples are passed through the decoder half of the VAE whose architecture mirrors that of the encoder, with MaxPool layers replaced by transposed 2D convolutions.

The sampled latent representation is then passed through the decoder function, $f_{de}$, which attempts to regenerate $X$ according to

$$f_{de} : z(X) \longrightarrow \hat{X}(z(X)). \quad (6)$$

The VAE's loss function has two main components, the first is the Kullback–Leibler (KL) divergence between the learned latent distribution $\mathcal{N}_\theta$ and a mean 0 unit variance normal distribution with diagonal covariance matrix according to

$$KL\left[\mathcal{N}_\theta, \mathcal{N}(\mathbf{0}, \mathbf{I})\right] = \frac{1}{2}\left(Tr(\Sigma) + z_\mu^T z_\mu - \log[\det(\Sigma)]\right). \quad (7)$$

This KL divergence regularizes the network's latent representation, pushing it towards a mean 0 unit variance distribution so that the latent space can be easily sampled and traversed without embeddings being placed in fragmented distant islands.

The second component of the loss function is the reconstruction loss defined as the negative log-likelihood of the re-generated images

$$R = -\log[Pr(X|z(X))], \quad (8)$$

which pushes the network towards accurately reconstructing the input images. The total VAE training loss is then defined as a weighted sum

$$C = R + wKL, \quad (9)$$

where the weight $w$ is typically chosen to be $< 1$ to allow the network to develop some structure in the latent embedding.

Despite using a highly compressed representation, the VAE is able to reconstruct highly accurate, very detailed images, as shown in Figures 6 and 7. The overall absolute percent error statistics for the training and test data of the VAE are shown in Figure 8.

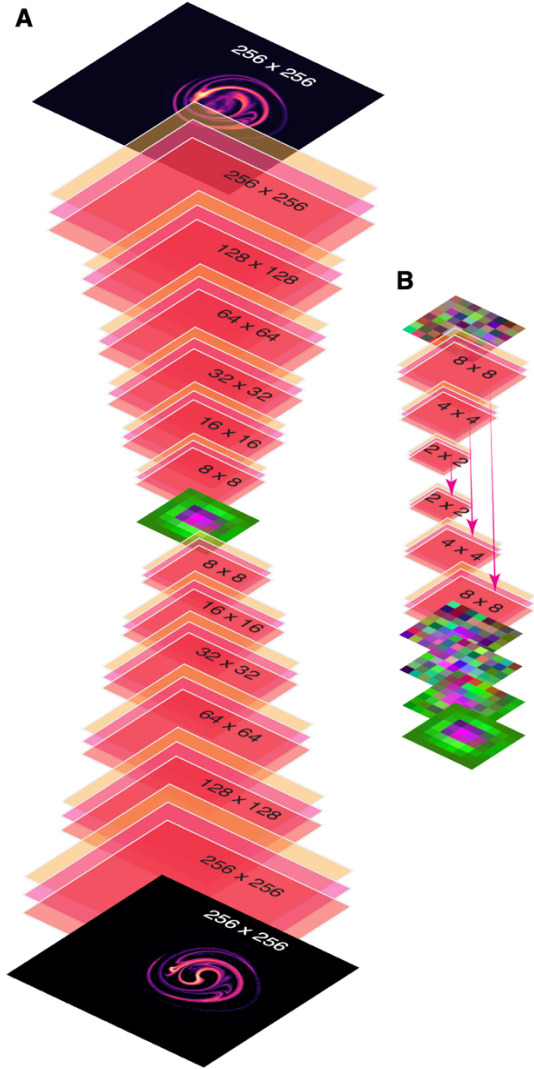

FIG. 5. **A:** High level overview of the VAE architecture (A) showing image size reduction from $256 \times 256$ pixels down to a $8 \times 8 \times 3$ latent embedding. **B:** The much smaller diffusion UNet working directly in the latent embedding space.

If we represent a $256 \times 256$ pixel input image as $X$, then the encoder function of the VAE, $f_{en}$ maps $X$ to mean $z_\mu$ and log variance $z_\Sigma$:

$$f_{en} : X \longrightarrow [z_\mu(X), z_\Sigma(X)], \quad (5)$$

which are then used to define a normal distribution from which latent representations $z$ are sampled

$$z(X) \sim z_\mu(X) + \Sigma(X)\mathcal{N}(\mathbf{0}, \mathbf{I}), \quad \Sigma(X) = \exp[z_\Sigma(X)/2].$$

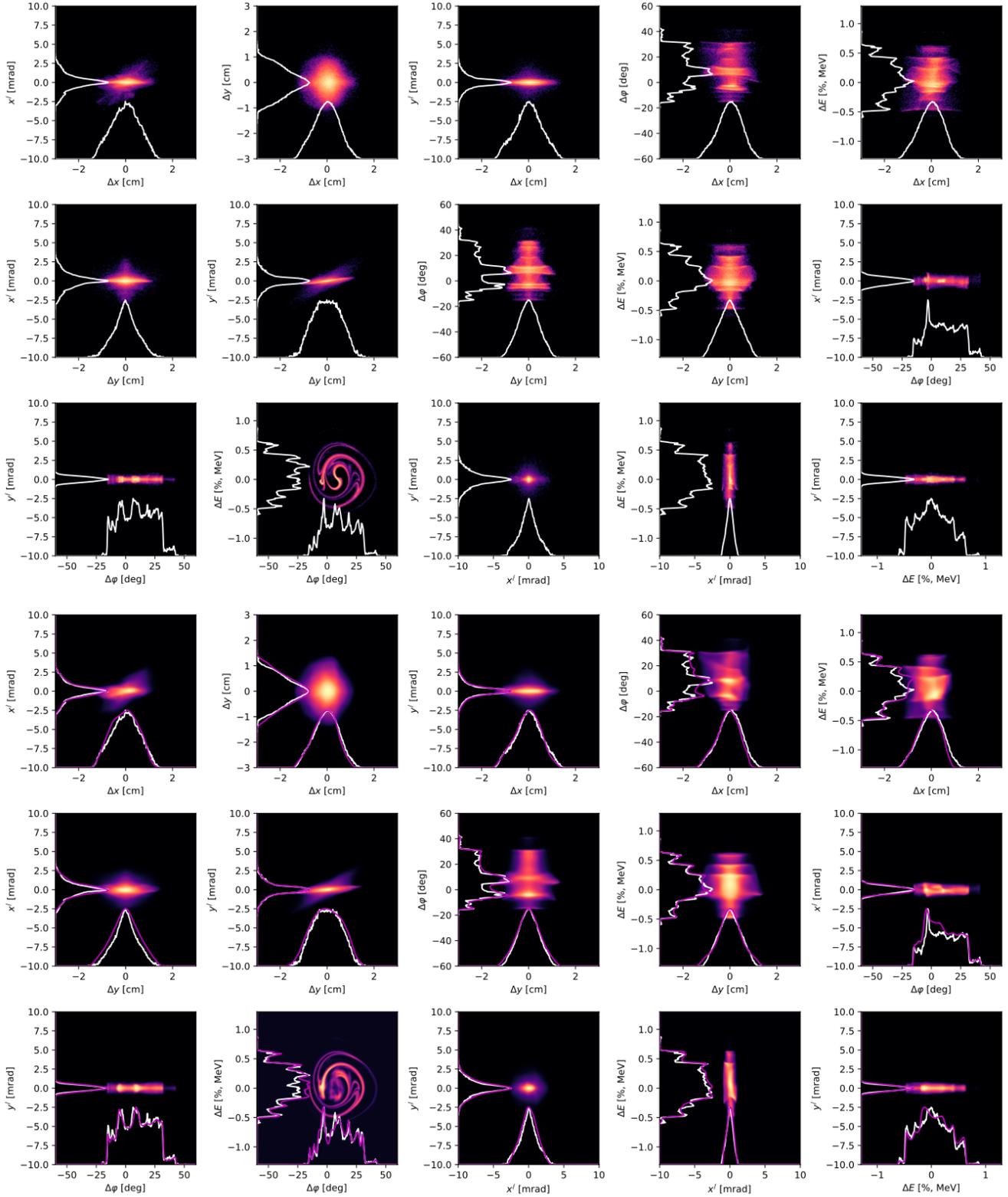

FIG. 6. The top 15 images show one example of the true phase space projections at module 48 for a test data point and the bottom 15 images show the generated images. The white curves are projections of the true data and the magenta curves are those of the generated data.

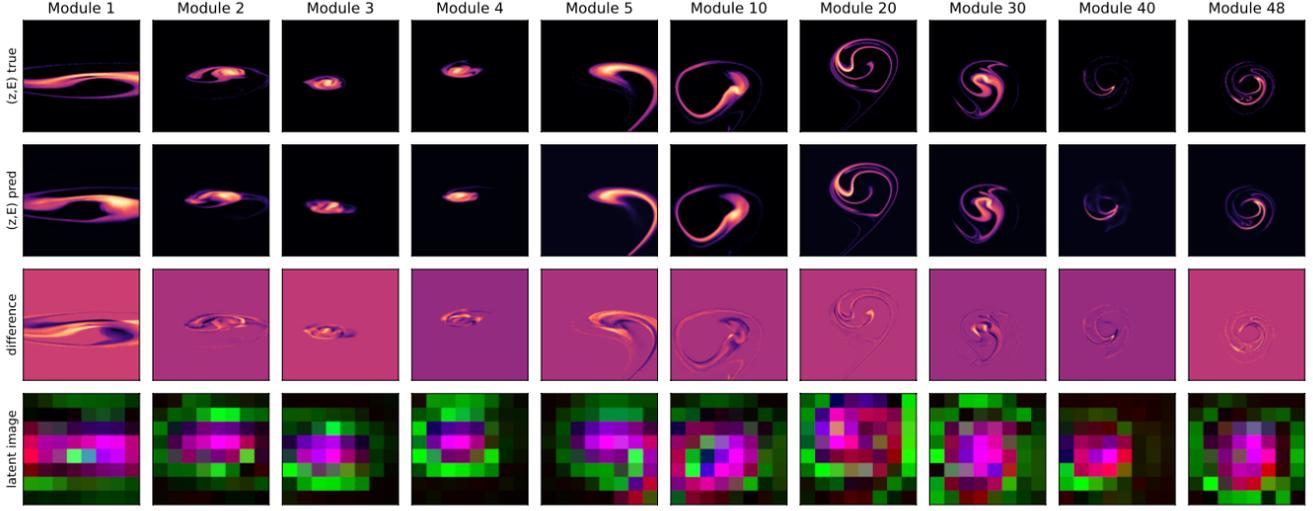

FIG. 7. The true and predicted $(\Delta\varphi, \Delta E)$ projections are shown at several modules along LANSCE along with the difference and the latent embeddings.

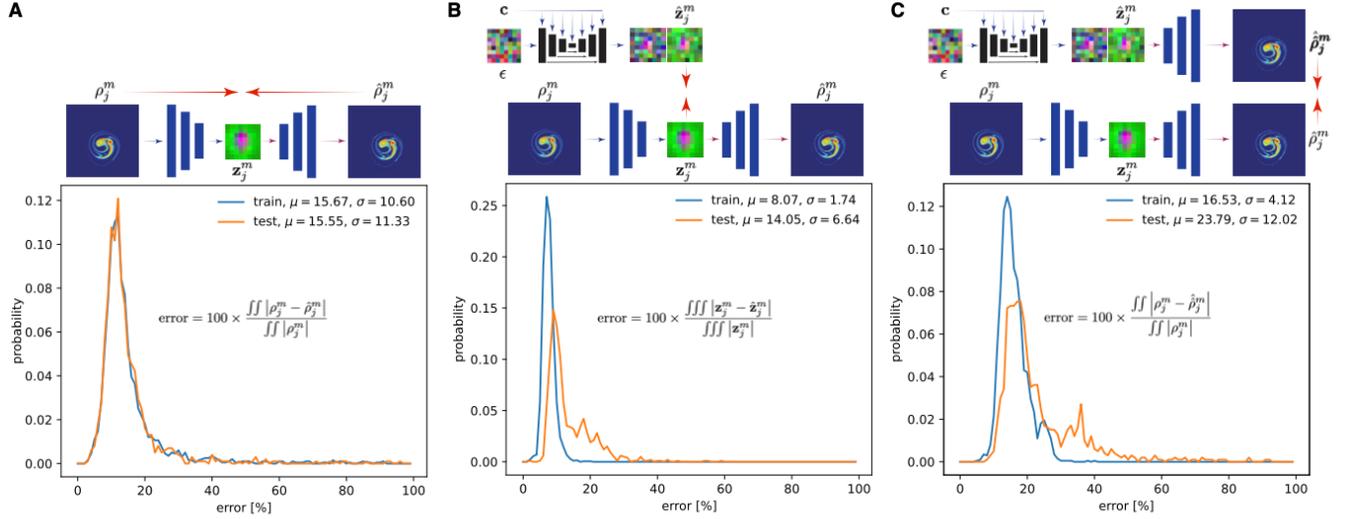

FIG. 8. Percent absolute error statistics for VAE reconstructions over the training and test data (A), for conditional diffusion reconstructions of the latent representations (B), and for the VAE decoder reconstructing the original images using the diffusion-based latent reconstructions (C).

## IV. Conditional Guided Latent Diffusion

In what follows, we sometimes drop the phase space coordinate arguments and refer to the projections above as

$$\rho_j^m(\mathbf{p}, \rho^0) \in \{\rho_1^m(\mathbf{p}, \rho^0), \rho_2^m(\mathbf{p}, \rho^0), \ldots, \rho_{15}^m(\mathbf{p}, \rho^0)\}, \quad (10)$$

to emphasize their dependence on the initial beam density $\rho^0$ and on the RF parameter settings

$$\mathbf{p} = [A_1, \theta_1, A_2, \theta_2, A_3, \theta_3, A_4, \theta_4], \quad (11)$$

which are the amplitude set points ($A_i$) and the phase set points ($\theta_i$) of the first 4 RF modules which make up the 201.25 MHz drift tube linac section of the LANSCE accelerator (all other RF settings are fixed).

We focus on this section of the accelerator because the beam's characteristics are most sensitive to these RF modules, having the greatest impact on the overall downstream performance. This is why in practice these are the first knobs that the operators tune in order to minimize beam loss or adjust other beam properties. For any given initial beam distribution $\rho^0$ and any RF settings p, we test whether we can reconstruct the 2D beam phase space projections from a vector of beam currents as measured at all 48 RF module locations

$$\mathbf{b} = [b_1, b_2, \ldots, b_{48}], \quad (12)$$

where the individual beam current measurements are as defined above in Equation 2. We assume that we do not have access to $\rho^0$ which in practice is a time-varying $\rho^0(t)$ which undergoes step changes with every source change. Our overall virtual diagnostic is then tasked with the mapping:

$$\mathbf{c} = [\mathbf{p}, \mathbf{b}, \mathbf{d}, m] \longrightarrow \rho_j^m, \quad (13)$$

where d is a vector which defines which of the 15 projections we would like to generate.

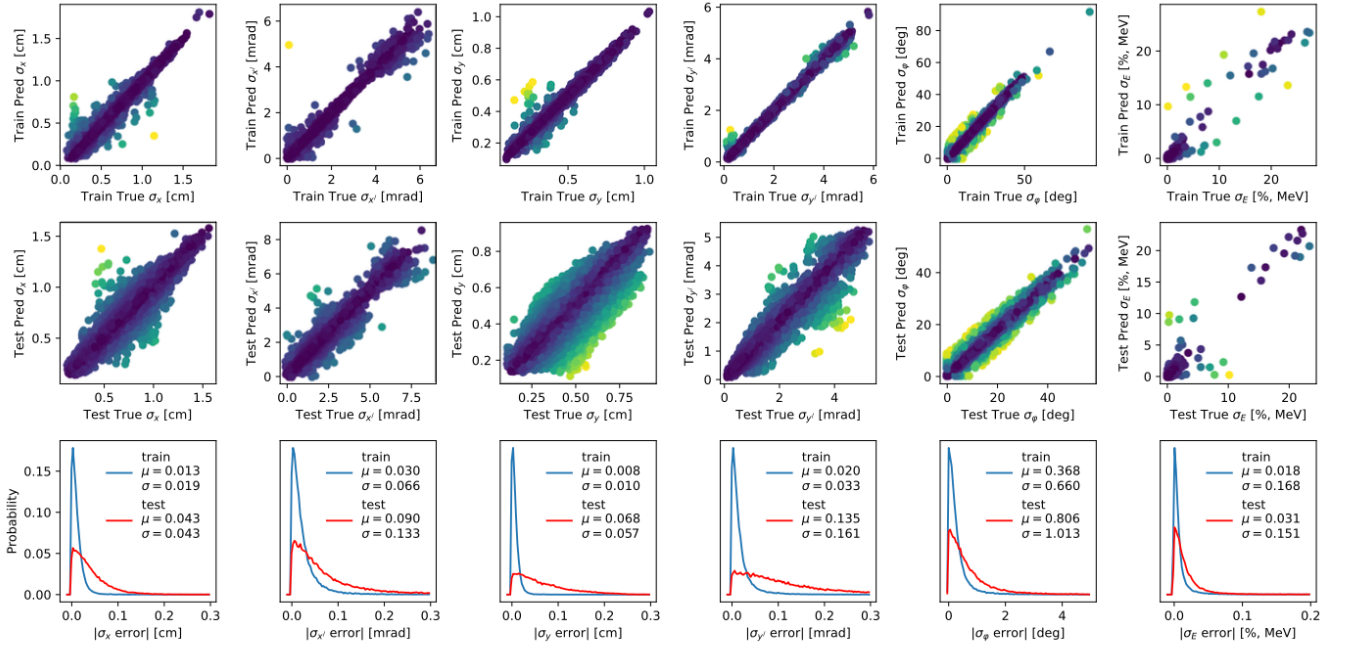

FIG. 9. Error statistics calculated by fitting Gaussians to each 1D projection of each image and then comparing the $\sigma$ values for both training and test data predictions. The top row shows true and predicted results for the training data, the middle row shows results for test data, and the bottom row shows histograms of the absolute errors.

As discussed in [27], a good physics-informed choice for **d** is to use a six dimensional vector whose positions correspond to the six phase space dimensions (x, y, $\varphi$, x/, y/, E) so that for example for $\rho^m_1$ (x, y), **d** = [1, 1, 0, 0, 0, 0] and when generating $\rho^m_{15}$ (y/, E), then **d** = [0, 0, 0, 0, 1, 1].

This physics-informed method is found to work better than simply choosing a scalar d ∈ {1/15, 2/15,…, 1}, which arbitrarily places certain projections closer to each other than others. For example, such an approach would condition the $\rho^m_1$(x,y) projection with 1/15, the $\rho^m_2$ (x,$\varphi$) projection as 2/15, and the $\rho^m_3$(x,x/) projection as 3/15, thus placing $\rho^m_2$(x,$\varphi$) twice as close as $\rho^m_3$(x, x/) to $\rho^m_1$(x, y), although all 3 projections share just the single x-axis. In contrast to this, for the physics-informed 6D vector approach conditions the projections $\rho^m_1$ (x, y), $\rho^m_2$ (x, $\varphi$), and $\rho^m_3$(x, x/) with the vectors $\mathbf{d}_1$ = [1,1,0,0,0,0], $\mathbf{d}_2$ = [1,0,1,0,0,0], and $\mathbf{d}_3$ = [1, 0, 0, 1, 0, 0], respectively, so that all of these projections that share the x-axis have the same pairwise inner products $\mathbf{d}_i \cdot \mathbf{d}_j$ = 1. Furthermore, completely independent projections, such as $\rho^m_{12}$($\varphi$,E) represented by $\mathbf{d}_{12}$ = [0, 0, 1, 0, 0, 1] and $\rho^m_{13}$(x/, y/) represented by $\mathbf{d}_{13}$ = [0, 0, 0, 1, 1, 0] are orthogonal with $\mathbf{d}_{12} \cdot \mathbf{d}_{13}$ = 0. This method also outperforms one-hot encoding d as 15-dimensional unit basis vectors of the form $\mathbf{d}_1$=[1,0,...,0], $\mathbf{d}_2$=[0,1,0,...,0], which forces each projection to be orthogonal, ignoring the shared axes.

Details of the diffusion process are available in the literature [20–25]. Here we provide only a quick overview. To generate latent images z, generative diffusion works by learning how to reverse a diffusion stochastic differential equation which can be described using a finite difference approximation as an iterative diffusion-based noising process which converts z into mean zero unit variance Gaussian noise according to

$$z_1 = \sqrt{1-\beta_1}z + \sqrt{\beta_1}\epsilon_1$$
$$z_2 = \sqrt{1-\beta_2}z_1 + \sqrt{\beta_2}\epsilon_2$$
$$\vdots$$
$$z_T = \sqrt{1-\beta_T}z_{T-1} + \sqrt{\beta_T}\epsilon_T, \quad \epsilon_k \sim \mathcal{N}(\mathbf{0}, \mathbf{I}). \quad (14)$$

The noise schedule $\{\beta_k, k \in 1,\ldots,T\}$ is chosen as an increasing sequence such that $0 < \beta_{k-1} < \beta_k < 1$ with typical starting and ending values of $\beta_1 = 1e-4$, $\beta_T = 1e-2$. For large high-resolution images (1024 × 1024) the sequence length is typically $T = 1000$, but for the much smaller latent images that we use for latent diffusion we can work with $T = 100$. This is one additional factor that greatly speeds up latent diffusion relative to standard image space diffusion, besides working with a smaller UNet we also need fewer diffusion steps.

For an image evolving according to Equation 14, a sample $z_k$ can be generated in a single shot from $z$ without having to go through the Markov chain one step at a time as it is easily calculated that

$$z_k = \sqrt{\alpha_t}z + \sqrt{1-\alpha_t}\epsilon, \quad \epsilon \sim \mathcal{N}(\mathbf{0}, \mathbf{I}), \quad (15)$$

where

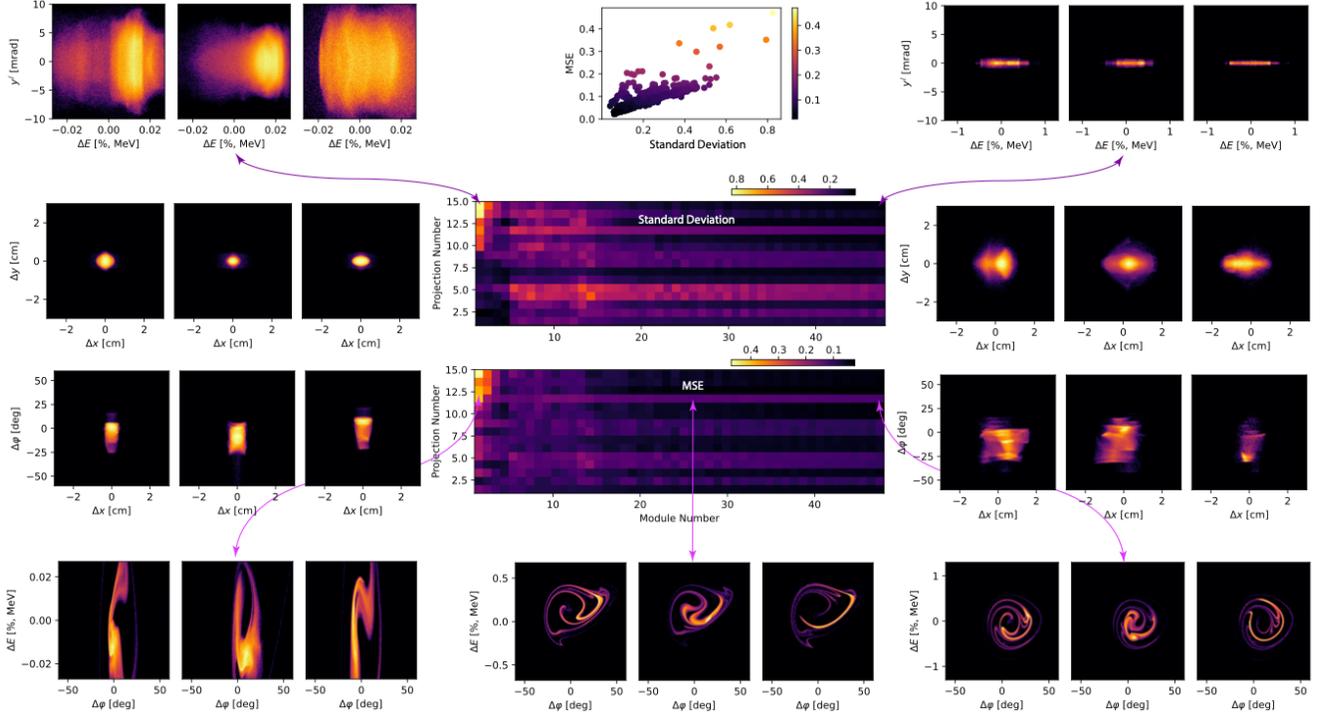

FIG. 10. Prediction error compared to variation of the data as a function of location in LANSCE (module number) and projection number. The top three figures in the middle column show that as expected MSE of the predictions rises with variation of the data (if there was no variation at all then the model could just memorize 1 image). In the first and third columns, three random examples of projections are shown at module 1 and also at the end of the accelerator at module 48, the bottom row shows three random longitudinal phase space projections of the beam including at the middle of the accelerator.

$$\alpha_k = \prod_{j=1}^{k}(1-\beta_j). \quad (16)$$

Equation 16 and $0 < \beta_j < 1$ implies that $\alpha_k$ converges towards 0 as $k$ increases and therefore from Equation 15 it is clear that $z_k$ approaches a sample from $\mathcal{N}(\mathbf{0}, \mathbf{I})$. Equation 15 makes it easy to sample the state of an image $z$ at an arbitrary diffusion time $k$. We can analytically approximate the conditional diffusion distribution of a reverse diffusion step of moving from $z_k$ to $z_{k-1}$ by a combination of Bayes' rule and a change of variables, as

$$q(z_{k-1}|z_k, z) = \mathcal{N}\left[\frac{1-\alpha_{k-1}}{1-\alpha_k}\sqrt{1-\beta_k}z_k + \frac{\sqrt{\alpha_{k-1}}\beta_k}{1-\alpha_k}z,\right.$$
$$\left.\frac{(1-\alpha_{k-1})\beta_k}{1-\alpha_k}\mathbf{I}\right]. \quad (17)$$

Having this analytic form (and making several additional simplifying assumptions as described in the literature) allows us to design the diffusion UNet to perform iterative de-noising steps by predicting and removing the noise that is added during the diffusion process $D_k$ which we model using a UNet $D_\theta$ according to

$$D_k[z_t, \mathbf{c}] = \frac{1}{\sqrt{1-\beta_k}}z_k - \frac{\beta_k}{\sqrt{1-\alpha_k}\sqrt{1-\beta_k}}D_\theta[z_t, t, \mathbf{c}]. \quad (18)$$

The overall architecture uses a U-net [34], with the setup that was developed for the PixelCNN [35], which uses group normalization instead of regular normalization [36], two residual CNN blocks at each resolution of $8 \times 8$, $4 \times 4$, and $2 \times 2$, and attention layers at the $4 \times 4$ and $2 \times 2$ pixel embedding levels [37–42].

Reconstruction error statistics for both the training and test data are shown in Figures 8, 9, and 10. In Figure 8 we can see a tradeoff of the latent diffusion approach as the overall test error grows along the encode, generate, decode sequence. Although it is faster than diffusion in image space, in the latent diffusion approach each generative model, and especially compression, adds its own error to the overall final result. In our case, and in most use cases, this tradeoff is acceptable if a large/powerful enough autoencoder is used. In Figure 9 we see a comparison between training and test data of the accuracy of the predicted first moments of the beam after it has been re-generated from its latent representation by the latent diffusion process. As expected performance is worse for test data than for training data, but overall the results are still highly accurate with $\sigma$ errors much smaller than the $\sigma$ values of the beams themselves. In Figure 10 we see a more detailed breakdown of the test error as the dependence of MSE test error is shown relative to module number and projection number. As expected we see that the method's performance is related to the variance of the data (if there was only 1 single beam image and zero variance then the model could just memorize that).

## V. ADAPTIVE LATENT DIFFUSION

ML for non-stationary systems is an open problem and an active field of research [43–46]. Typical approaches for non-stationary systems rely on detecting significant changes after which the weights of neural networks are updated/re-trained with new information or are continuously trained to keep up with continuous changes [47–49]. Approaches have been developed for the case of covariate shift, where the input distribution $P(\mathbf{x})$ is different for training and test data, but the conditional distribution of output values $P(y|\mathbf{x})$ remains unchanged [50], based on importance-weighting (IW) techniques [51]. IW methods have also been developed using kernel mean matching methods [52] and by minimizing the Kullback-Leibler divergence between a test data density distribution and its estimate [53, 54]. Methods have also been developed for extracting frequencies and amplitudes from time-series data [55], and Bayesian methods have been used with sinusoidal kernels for periodic time-varying systems [56].

Recently an adaptive machine learning approach has been developed which combines deep learning with the state-of-the-art robust adaptive feedback mechanism from control theory, known as bounded extremum seeking, for stabilizing and optimizing analytically unknown dynamic systems [57–61]. This approach was first demonstrated for automatic control of the $(z, E)$ longitudinal phase space of the LCLS electron beam [10], it has been incorporated within the architecture of generative models for real-time adaptive latent space tuning for real-time tracking of time-varying acceleartor beams based on limited measurements [13–15], for tracking time-varying 3D electron densities in coherent diffraction imaging [62], and it has been used for adaptive tuning of conditional vectors that guide generative diffusion models as virtual particle accelerator beam diagnostics [27, 28].

For our adaptive application, we consider the case where we have access to a 1D projection of one of the 2D images that we are trying to generate. For example, the projection of the longitudinal phase space distribution along the phase or time axis, as defined by

$$\rho^m(\varphi) = P_E \rho^m(\varphi, E) = \int_E \rho^m(\varphi, E) dE, \quad (19)$$

is a measure of the bunch current profile at module $m$. Sometimes the beam can also be projected along the energy axis, as defined by

$$\rho^m(E) = P_\varphi \rho^m(\varphi, E) = \int_\varphi \rho^m(\varphi, E) d\varphi. \quad (20)$$

Such a measurement is performed by passing the beam through a dispersive element, such as a dipole bending magnet, thus correlating transverse position with energy, and then impacting a phosphor screen or scintillator. For any such projection, $P$, of a 2D beam phase space density $X(\tau)$ we can simulate the same projection, $\hat{P}$, and apply it to the diffusion-generated version of the 2D distribution $\hat{X}(\tau)$. Comparing the measured and simulated projections we can define a simple cost function

$$C(\tau, \mathbf{c}) = \left\| \hat{P}\hat{X}(\mathbf{c}, \tau) - PX(\tau) \right\|$$
$$= \int \left| \hat{P}\hat{X}(\mathbf{c}, \tau) - PX(\tau) \right|, \quad (21)$$

where we have highlighted the fact that the generated projection $\hat{X}$ depends on the conditional vector $\mathbf{c}$. Our goal is then to apply bounded extremum seeking to adaptively tune the conditional vector $\mathbf{c}$ to continuously track a time-varying $PX(\tau)$.

For this kind of time-varying optimization problem, the bounded extremum seeking method, as first described in [58], simply follows the dynamics

$$\dot{c}_j = \sqrt{\alpha \omega_j} \cos\left[\omega_j \tau + kC(\tau, \mathbf{c})\right], \quad (22)$$

where $\omega_i \neq \omega_j$ are distinct dithering frequencies. The term $\alpha$ defines the steady-state oscillation amplitudes of the tuned parameters. Once a minimum of $C(\tau, \mathbf{c})$ is reached, $c_j(\tau)$ will oscillate about its minimizing value $c_j^*(\tau)$ with a trajectory of the form

$$c_j(\tau) \approx c_j^*(\tau) + \sqrt{\frac{\alpha}{\omega_j}} \sin(\omega_j \tau + \theta),$$

where $\theta$ is some phase offset. To get close to the minimum and for the adaptive dynamics to be faster than the system's time-variation, the $\omega_j$ must be sufficiently large relative to $\|\partial C/\partial t\|$. The term $\alpha$ also acts as a feedback gain together with the term $k$ as on average the system dynamics of Equation 22 can be approximated arbitrarily accurately by the gradient flow dynamics

$$\dot{\mathbf{c}} = -\frac{k\alpha}{2} \nabla_\mathbf{c} C(\tau, \mathbf{c}), \quad (23)$$

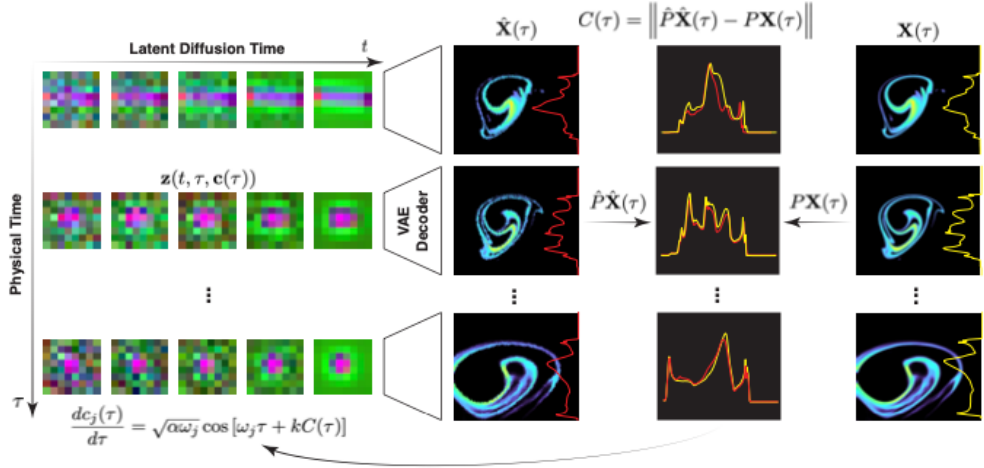

FIG. 11. Overview of adaptive conditional tuning for a virtual 2D diagnostic that can track a time-varying beam based on 1D projections. In this figure $\tau$ represents physical time relative to which accelerator components and initial conditions of the beam are changing while $t$ represents generative diffusion process time. If a 1D projection $P\mathbf{X}(\tau)$ of one of the beam's 2D distributions is available for measurement it can be compared to a simulated projection $\hat{P}$ of the generated representation of the same 2D distribution given as $\hat{P}\hat{\mathbf{X}}(\tau)$. A cost function defined by the difference between these projections is then used to adjust the conditional vector $\mathbf{c}(\tau)$ to adaptively track the time-varying beam.

which for $k\alpha > 0$ finds and tracks the minimum of the analytically unknown function $C(\tau, \mathbf{c})$. Note that on-average the dynamics are a gradient flow, the adaptive feedback as defined in Equation 22 is not attempting any kind of discrete gradient estimation, which makes the method incredibly robust for high-dimensional and noisy dynamic systems. The method has been demonstrated in-hardware for automatic particle accelerator tuning including automatic FEL output power maximization at both the LCLS and the EuXFEL [63], for real-time multiobjective optimization of the AWAKE electron beamline [64], and for automatic beam loss minimization at LANSCE [65].

An overview of the adaptive setup is shown in Figure 11, which emphasizes that there are two time scales to consider. The time $\tau$ is related to physical time relative to which the beam and accelerator are changing and evolving and the conditional vector $\mathbf{c}(\tau)$ must be adjusted to track the time-varying object. The second time scale $t$ is the diffusion process time representing the diffusion-based image generation process.

While this process can in principal be applied for any beam projection at any accelerator location, in this demonstration we focus on using the 1D energy projection as our measurement. We use the energy projection because we have locations in LANSCE where the beam goes around a bend and we can measure the energy spread on a phosphor. We focus on modules near the high energy end of LANSCE because that is where the energy screens are located and that is also where we need the finest control over the beam's $(\varphi, E)$ phase space because high energy beam loss contributes most to equipment damage and radioactive activation of accelerator components, and also the final beam energy must be finely controlled for experiments. Figure 12 shows three examples of an initial beam distribution which then changes because of drifting RF parameters, where we assume that we do not have access to the true RF signals, but we are nevertheless able to track the time-varying beam distribution by continuously adaptively tuning the conditional vector of the generative diffusion process based only on the 1D energy projections. We choose the RF parameters to vary because during LANSCE operations the RF systems are the fastest time-varying signals and the RF pickup loops have arbitrary temperature-dependent offsets.

Because our adaptive feedback only has access to a 1D projection of a 2D image, the tracked reconstruction is not perfect as even an exact match of a 1D projection can come from uncountably many different 2D objects. For example, a square and a circle whose diameter is the same length as the side of the square can have the same 1D projection if their densities are defined correctly. Nevertheless, by starting at the correct solution, we see that we can accurately track the time-varying image (only missing minor interior details) based only on tracking the 1D projection because the beam is not just an arbitrary shape, rather it is defined as a function of RF parameter settings according to a very specific physical process and the diffusion model has learned how to accurately move around in this parameter space.

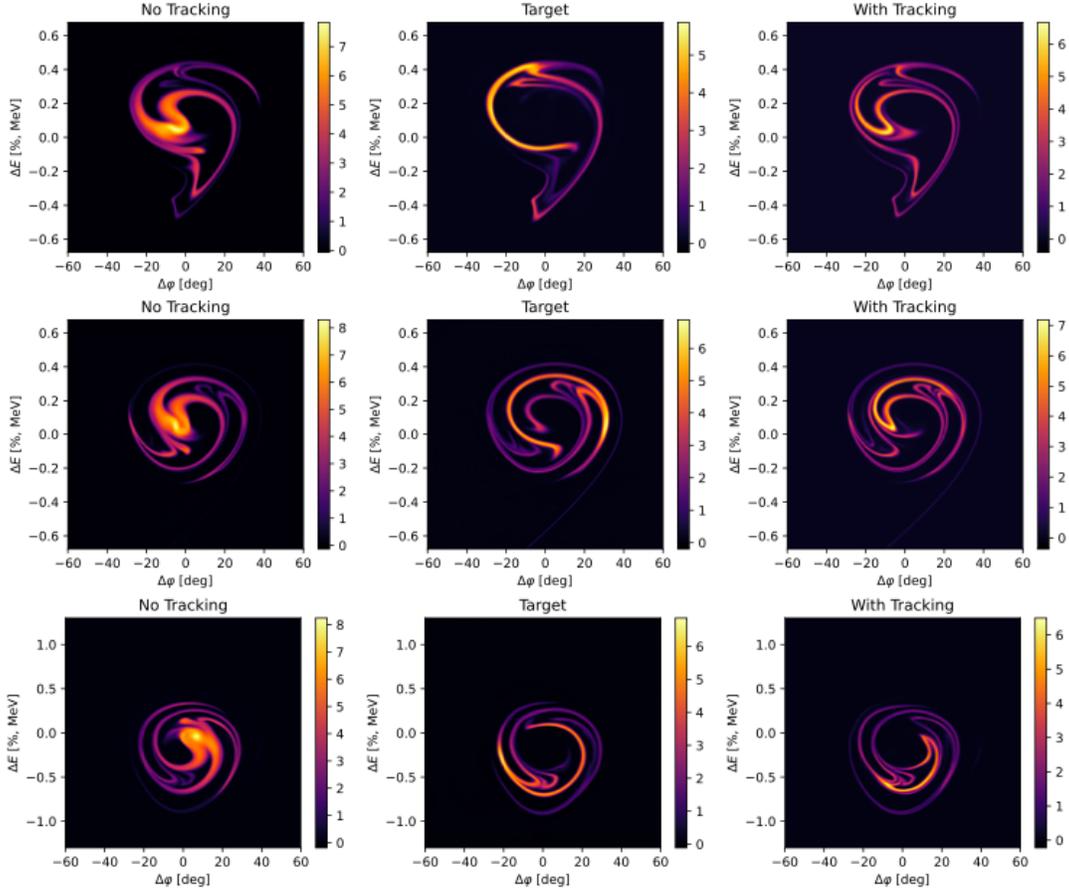

FIG. 12. Demonstration of adaptive longitudinal phase space tracking at LANSCE. The left columns show initial beam distributions, the middle columns show the distributions that the beams evolve to due to time-varying RF module settings, and the right columns show the adaptively tracked predictions. The top row is for module 28, the next is module 38, and the bottom row is module 48.

## VI. CONCLUSIONS

We have developed a conditionally guided latent diffusion model that can map waveforms of beam loss or beam current measurements along an accelerator to detailed 2D projections of the beam's 6D phase space density. We have also shown that if calibrated then adaptive condition tuning can track a time-varying beam based only on 1D projections even when the RF parameters begin to change and are unknown. This transformational method can be used at any particle accelerator to transform simple non-invasive devices into detailed beam phase space diagnostics. We demonstrate this concept via multi-particle simulations of the high intensity beam in the kilometer-long LANSCE linear proton accelerator. Future work will focus on expanding to a higher dimensional parameter study including additional RF cavities, magnets, and experimental demonstration.

## ACKNOWLEDGMENTS

This work was funded by the US Department of Energy (DOE) Los Alamos National Laboratory LDRD Program Directed Research (DR) project 20220074DR.